\documentclass{PoS}

\title{D-branes and Coherent Topological Charge Structure in QCD}

\ShortTitle{D-branes, Wilson bags, and Coherent Topological Charge Structure in QCD}

\author{\speaker{H. B. Thacker}\\
        University of Virginia, Charlottesville, VA 22904\\
        E-mail: \email{hbt8r@virginia.edu}}

\abstract{Monte Carlo studies of pure glue $SU(3)$ gauge theory using the overlap-based topological
charge operator have revealed a laminar structure in the QCD vacuum consisting of extended, 
thin, coherent, locally 3-dimensional sheets of topological charge embedded in 4D space, 
with opposite sign sheets interleaved. Studies of localization properties of Dirac 
eigenmodes have also shown evidence for the delocalization of low-lying modes on effectively 3-dimensional surfaces. In this talk, I review some theoretical ideas which suggest the possibility of
3-dimensionally coherent topological charge structure in 4-dimensional gauge theory and
provide a possible interpretation of the observed structure. I begin with Luscher's ``Wilson
bag'' integral over the 3-index Chern-Simons tensor. The analogy with a Wilson loop
as a charged world line in 2-dimensional
$CP^{N-1}$ sigma models suggests that the Wilson bag surface represents the world volume of a physical 
membrane. The large-N chiral Lagrangian arguments of Witten also indicate the existence of 
multiple ``k-vacuum'' states with discontinuous transitions between k-vacua at $\theta=$ odd multiples
of $\pi$. The domain walls between these vacua have the properties of a Wilson bag surface.
Finally, I review the AdS/CFT duality view of $\theta$ dependence in QCD. The dual realtionship between
topological charge in gauge theory and Ramond-Ramond charge in type IIA string theory suggests that
the coherent topological charge sheets observed on the lattice are the holographic image of
wrapped D6 branes.}

\FullConference{XXIVth International Symposium on Lattice Field Theory\\
                July 23-28, 2006\\
                Tucson, Arizona, USA}

\begin{document}

\section{Introduction}

One of the most important byproducts of the chiral lattice fermion revolution of
the late 1990's was a new definition of gauge field topological charge density $q(x)=
(g^2/16\pi^2)TrF\tilde{F}$ on the lattice.
This definition is constructed from an exactly chiral Dirac operator $D$ satisfying Ginsparg-Wilson relations.
The local pseudoscalar operator given by \cite{Hasenfratz97}
\begin{equation}
q_o(x) = \frac{1}{2}Tr\gamma_5 D(x,x)
\end{equation}
(here the trace is over both color and spin)
reduces to $q(x)$ in the continuum limit, and is in many respects a superior definition of topological
charge density compared to any ultralocal operator constructed directly from gauge links.
Recent studies of topological charge using the overlap-based \cite{Neuberger} topological charge density have
revealed a type of long-range structure that is profoundly different from what might have been
expected in an instanton-based
model of the QCD vacuum. The first such study \cite{Horvath03} produced the surprising result
that the $q(x)$ distribution
in a typical Monte Carlo gauge configuration is dominated by extended, coherent, thin 3-dimensional sheets
of topological charge. In each configuration, sheets of opposite sign are juxtaposed and are everywhere close together 
in what can roughly be described as a dipole layer which is spread throughout the 4-dimensional Euclidean
space (with various folds and wrinkles). The vacuum is thus permeated with what is locally a laminar structure
consisting of alternating sign sheets or membranes of topological charge. The thickness of these membranes is
typically a few lattice spacings, independent of the physical mass scale, and thus the membranes apparently
become infinitely thin in the continuum limit. This kind of ``subdimensional'' ordering, 
where coherence takes place on manifolds of lower dimensionality than spacetime itself, is closely
related to the appearance of contact terms in the two-point topological charge correlator.
In the continuum, the correlator $G(x) =\langle q(x)q(0)\rangle$ {\it is required by spectral considerations to be negative
for any nonzero separation} $|x|>0$. In practice (i.e. in Monte Carlo calculations), this requirement
places severe restrictions on what type of topological charge fluctuations can be dominant. 
For example, the negativity of the correlator rules out the dominance of bulk-coherent lumps of topological charge (e.g. finite
size instantons), since this would lead to a positive correlator over distances smaller than the instanton
size. Since the topological susceptibility can be obtained by integrating the 2-point correlator 
over all $x$, a positive susceptibility can only arise from a delta-function contact term at the
origin. The observed arrangement of thin, nearby layers of $q(x)$ with opposite sign builds up
a positive contact term at $x=0$ while maintaining the required negativity of the correlator for finite separation. 
In recent overlap-based studies, the only models that have been found to be 
dominated by instantons are the $CP^1$ and $CP^2$ sigma models \cite{Lian06}. But those models are dominated by
{\it small} instantons with a radius of order lattice spacing, which goes to zero in the continuum limit.
Because of this, the instantons contribute to the positive delta-function contact term, but 
do not contribute at all to the finite $x$ correlator.   

\begin{figure}
   \begin{center}
     \vskip -0.15in
     \centerline{
     \includegraphics[width=9.5truecm,angle=-90]{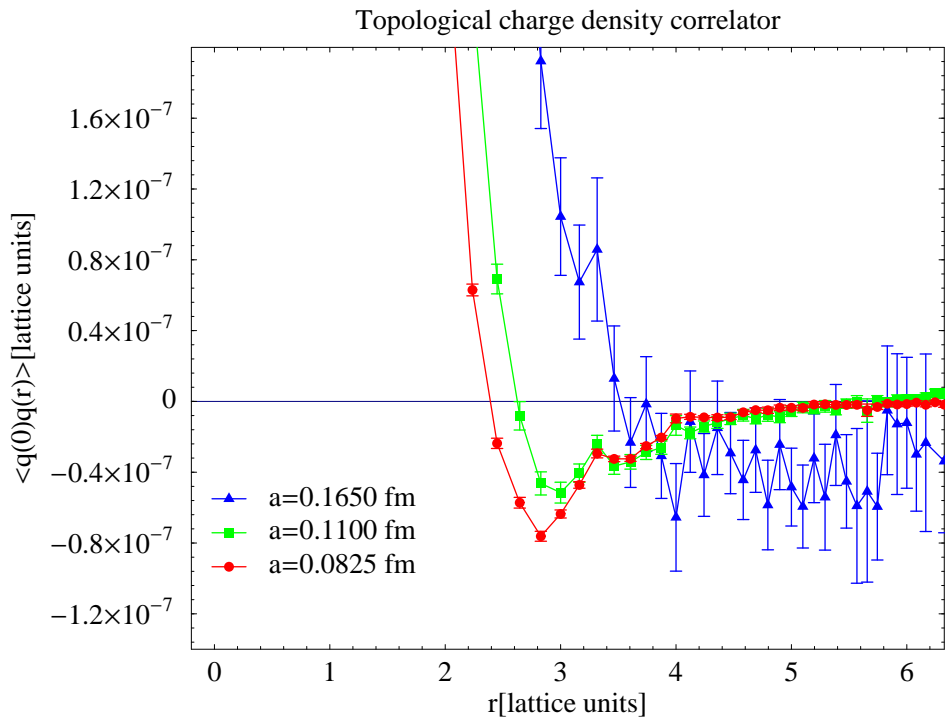}
     }
     \vskip 0.15in
     \caption{2-point functions of the overlap based topological charge correlator in pure glue QCD.}
     \label{fig:qcd_2point}
   \end{center}
\end{figure}
The presence of coherent sheets of topological charge is responsible for the positive
contact term in the correlator at $x=0$. The range of this contact term on the lattice is associated with the
thickness of the sheets, both being a few lattice spacings and 
approaching zero in physical units. Fig. \ref{fig:qcd_2point} shows the topological
charge correlator for pure glue SU(3) gauge theory at several values of lattice spacing \cite{Horvath_corr}. 
In addition to the large positive contact term appearing for $r\equiv|x|\leq 2$, 
the other prominent feature is a clear negative dip at a few sites separation. At a calculational level, this negative dip arises from the nearby juxtaposition of positive and negative topological charge layers.

\begin{figure}
   \begin{center}
     \vskip -0.15in
     \centerline{
     \includegraphics[width=9.5truecm,angle=-90]{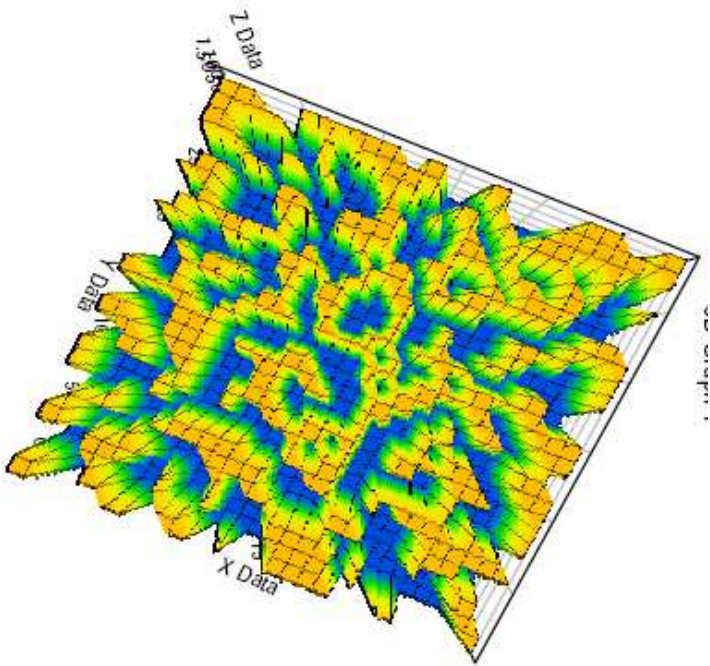}
     }
     \vskip 0.15in
     \caption{Positive and negative regions of topological charge for a typical $CP^3$ configuration.}
     \label{fig:labyrinth}
   \end{center}
\end{figure}

Recent calculations have provided further evidence for the existence of 
these extended 3-dimensional sheets of topological charge \cite{Ilgenfritz}. 
The study of localization properties of low Dirac eigenmodes in Monte Carlo configurations has also shown that 
low-lying modes are delocalized along effectively 3-dimensional surfaces in 4D space.\cite{MILC} 
This is just what
one would expect for low Dirac eigenmodes in the presence of 3-dimensional coherent topological 
charge sheets. Long ago, Diakonov
and Petrov \cite{Diakonov} pointed out that topological charge fluctuations could lead to a chiral condensate. In the context
of the instanton liquid model, the finite density of near-zero eigenmodes required to produce a condensate is
generated by approximate 'tHooft zero modes associated with the instantons. But, more generally, coherent regions 
of positive (negative) topological charge will attract left (right) chiral Dirac modes, and produce low-lying states 
in which the Dirac mode is bound to the topological charge fluctuation. A rather natural picture for 
Goldstone boson propagation emerges, in which quarks in a pion ``skate'' along the surface of a coherent topological charge
membrane via Dirac modes which are delocalized along this surface. Further studies of low Dirac eigenmodes  and 
their relationship to the topological charge distribution should help to clarify this picture.

Since the initial discovery of coherent topological charge sheets in QCD, similar methods have been
applied to the study of 2-dimensional $CP^{N-1}$ sigma models \cite{Ahmad05,Lian06}. For $N>3$, the
topological charge distribution was found to be dominated by thin 1-dimensionally coherent membrane-like
structures with interleaved membranes of opposite sign. 
Fig. \ref{fig:labyrinth} shows a typical
overlap-based topological charge distribution for $CP^3$, where $sign(q(x))$ is plotted. 
As I will discuss, this interleaved arrangement of approximately 1-dimensional
coherent regions is exactly what one would expect
as the analog of the 3-dimensional structures in 4D gauge theory. In both cases, 
the coherent structure has codimension 1, i.e. the dimensionality of a domain wall.  

\section{Wilson Loops, Wilson bags, and Chern-Simons tensors}

Though there is now considerable numerical evidence for the existence of 3-dimensionally coherent 
structure in 4D gauge configurations, the theoretical significance of this structure is far from clear. 
In this talk I will review a number of related
theoretical developments which lead to a plausible interpretation of the observed structure. In fact, I will argue
that the Monte Carlo observations essentially support and clarify a view of topological charge in QCD which was suggested long
ago by work of Luscher \cite{Luscher78} and Witten \cite{Witten79,Witten80} and has re-emerged more recently
in the context of AdS/CFT string/gauge duality. 

Luscher's discussion begins with the fact that nonzero topological
susceptibility implies the presence of a massless pole in the two-point correlator of the Chern-Simons current.
Let us define the {\it abelian} 3-index Chern-Simons tensor
\begin{equation}
\label{eq:CStensor}
A_{\mu\nu\rho} = -Tr\left(B_{\mu}B_{\nu}B_{\rho}+\frac{3}{2}B_{[\mu}\partial_{\nu}B_{\rho]}\right)
\end{equation}
where $B_{\mu}$ is the Yang-Mills gauge potential. We consider the Chern-Simons current that is dual to this tensor,
\begin{equation}
\label{eq:CS4d}
j_{\mu}^{CS} = \frac{1}{32\pi^2}\epsilon_{\mu\nu\rho\sigma}A_{\nu\rho\sigma}\,\,.
\end{equation}
Although $j_{\mu}^{CS}$ is not gauge invariant, its divergence is the gauge invariant topological charge density
\begin{equation}
\label{eq:csdiv}
\partial_{\mu}j_{\mu}^{CS} = \frac{1}{16\pi^2}Tr F\tilde{F} = q(x) \,\,.
\end{equation}
Choosing a covariant gauge, $\partial_{\mu}A_{\mu\nu\rho}=0$, the correlator of two Chern-Simons currents
has the form
\begin{equation}
\label{eq:cscorr}
\langle j_{\mu}^{CS}(x)j_{\nu}^{CS}(0)\rangle = \int \frac{d^4q}{(2\pi)^4}\; e^{-iq\cdot x}\; \frac{q_{\mu}q_{\nu}}{q^2} G(q^2) \,\,.
\end{equation}
From (\ref{eq:csdiv}) we see that $G(q^2)$ must have a $q^2=0$ pole whose residue is the topological susceptibility,
\begin{equation}
G(q^2) \sim \frac{\chi_t}{q^2} \,\,.
\end{equation}
This long-range correlation constitutes a ``secret long-range order'' of
gauge fields associated with their topological charge fluctuations. Since the CS
current is not gauge invariant, the presence of a $q^2=0$ pole does not imply the existence of a massless particle
(and pure-glue QCD certainly doesn't have one).
On the other hand, the pole has a gauge invariant residue ($\propto\chi_t$) and cannot be transformed away.
So it characterizes a physically significant long range coherence in the gauge field associated with 
topological charge fluctuations.

Luscher's analysis of QCD topological structure in terms of Wilson bags can be understood as a generalization of the analysis 
of similar properties in the 2-dimensional $CP^{N-1}$ sigma models.
These models provide a quite detailed 2D analog of the coherent structure observe in 4D QCD. 
The $CP^{N-1}$ models have a
$U(1)$ gauge invariance and have classical instanton solutions which come in all sizes. (Just like pure-glue QCD these models 
are classically scale invariant and acquire a mass scale via a conformal anomaly.) 
In the continuum, the Euclidean action is
\begin{equation}
{\cal L} = (D_{\mu}z_i)^*(D_{\mu}z_i)
\end{equation}
where $D_{\mu}=\partial_{\mu}+ iA_{\mu}$, and 
$z_i,\; i=1,\ldots,N$ is an $N$-component complex vector constrained to $z^*_iz_i=1$.
In 2D U(1) gauge theories like $CP^{N-1}$,
the topological charge density in the continuum is just $(1/2\pi)\epsilon_{\mu\nu}\partial_{\mu}A_{\nu}$ and the 
Chern-Simons current is just the dual of the gauge potential,
\begin{equation}
\label{eq:CS2d}
j_{\mu}^{CS}= \frac{1}{2\pi}\epsilon_{\mu\nu}A_{\nu}
\end{equation}
Just as in 4D QCD, nonzero topological susceptibility implies the presence of a $q^2=0$ pole in the CS current correlator.
But in the two-dimensional case, this same pole appears in the $A_{\mu}$ correlator and
is responsible for confinement of $U(1)$ charge
via a linear coulomb potential. In
the $CP^{N-1}$ models, the gauge field $A_{\mu}$  
is an auxiliary field which has no kinetic term in the action. It's equation
of motion sets it equal to the $U(1)$ current of the matter fields,
\begin{equation}
\label{eq:CFI}
A_{\mu} = i\left(z_i^*\partial_{\mu}z_i-(\partial_{\mu}z_i)^*z_i\right)
\end{equation}
However, as can be shown explicitly in the large-N approximation, the quantum effect of closed $z$-loops 
produces a dynamically generated kinetic term $\propto F_{\mu\nu}^2$ in the low-energy effective action of the
gauge field. This produces a $q^2=0$ pole in the gauge field correlator which gives rise to a 
linear, confining coulomb potential between test charges, an area law for
fractionally charged Wilson loops, and nonzero topological susceptibility. 

Thus in 2-dimensional $U(1)$ gauge theories, topological susceptibility and confinement of $U(1)$ charge are equivalent phenomena. An instructive way to illustrate this is to introduce a nonzero $\theta$ term 
in the action over a two-volume $V$ enclosed by a boundary 
$C=\partial V$ with $\theta=0$ outside the boundary. After integration by parts, the theta term 
is equivalent to a Wilson loop around the boundary carrying a charge $\theta/2\pi$:
\begin{equation}
\label{eq:thetaterm}
\exp\left[\frac{i}{2\pi} \int d^2x \theta(x)\epsilon_{\mu\nu}F^{\mu\nu}\right] = \exp\left[\frac{i\theta}{2\pi}\oint_{C} A\cdot dx\right]
\end{equation}
If the topological susceptibility is nonzero 
\begin{equation}
\chi_t=\frac{\partial^2E(\theta)}{\partial \theta^2}|_{\theta=0} >0
\end{equation}
then for small nonzero $\theta$, the vacuum energy density $E(\theta)$ inside the loop will be greater than that
outside the loop, so the Wilson loop  will obey an area law,
\begin{equation}
\langle W(C)\rangle \propto \exp\left[-\left(E(\theta)-E(0)\right)V\right]
\end{equation}
In a Hamiltonian framework, the Wilson loop around the boundary corresponds to applying a background electric field $\theta$ by putting
opposite charges at either end of the 1-dimensional spatial box. The topological susceptibility is just 
the vacuum polarizability with respect to this field. This is essentially Coleman's original interpretation of $\theta$-dependence
in the massive Schwinger model \cite{Coleman76}. This general picture carries over to the $CP^{N-1}$ models 
(except for $CP^1$ and $CP^2$ \cite{Lian06}). 
The physics is somewhat different in the $CP^{N-1}$ models than it is in the massive Schwinger model because of the fact that $A_{\mu}$ is an auxiliary
field without a kinetic term. The confining electric flux tube in the $CP^{N-1}$ case actually represents the polarization of 
$z$ pairs in the vacuum between the test charges. 

On general principles, we expect the energy density $E(\theta)$ of the true vacuum to be periodic in $\theta
\rightarrow \theta+2\pi$. However, this periodicity can arise in two very distinct ways: analytically
or nonanalytically. Dilute instanton calculations produce an $E(\theta)$ which is smooth and periodic (polynomial
in $\cos\theta$). On the other hand, the mechanism described by Coleman in the massive 
Schwinger model for periodicity as a function of $\theta$ involves a 
discontinuous ``string breaking'' at $\theta=\pi$. For $\theta=\pi$, the original vacuum, with $\frac{1}{2}$ a unit of background flux
to the right, becomes degenerate with the one with $\frac{1}{2}$ unit of flux to the left. For $\theta>\pi$ it becomes 
energetically favorable to pop a charged pair out of the vacuum and screen off one unit of flux. The true ground
state shifts to the ``k-vacuum'' with $k=-1$ (where the local value of $\theta$ differs from the applied field
by minus one unit of flux.). In two dimensions, a charged particle world line can be regarded as a domain
wall between two different vacuum states. 
As $\theta$ goes from $\pi$ to $2\pi$, the vacuum energy inside the bag decreases.
At $\theta=2\pi$, the energy inside and outside are equal, the area law vanishes,
and the Wilson loop around $C$ is completely screened by the polarization of the vacuum inside $C$.
By Eq. (\ref{eq:CFI}) the screened unit-charged Wilson loop around $C$ is a filamentary thread of current,
similar to an edge current in a 2-dimensional superconductor. 
(In Lorentz gauge, $\partial_{\mu}A_{\mu}=0$, this current is conserved.) Since the $q(x)$ distribution is
just the curl of the $A_{\mu}$ field, the topological 
charge distribution associated with a Wilson
line excitation is a dipole layer consisting of two opposite sign layers of 
one-dimensionally coherent topological charge
membranes on either side of the Wilson line. This is just the type of structure that is seen in the Monte Carlo configurations.

In specifying the analogy between 2D U(1) theories and 4D SU(N) gauge theories, 
we take the Chern-Simons currents (\ref{eq:CS4d}) and (\ref{eq:CS2d}) to be directly
analogous. This means that the gauge field $A_{\mu}$
in the 2D theory should be identified {\it not} with the 4-dimensional gauge field, but with the abelian 3-index Chern-Simons
tensor (\ref{eq:CStensor}).
Like the gauge field $A_{\mu}$ in 2 dimensions, this is dual to the Chern-Simons current (\ref{eq:CS4d}).
Similarly, the Wilson loop or line excitations in the 2-dimensional $U(1)$ models correspond not to Wilson loops in 4D, 
but to ``Wilson bags,'' i.e. integrals of the Chern-Simons tensor over a 3-surface $\Sigma$.
\begin{equation}
\label{eq:WB}
B(\Sigma) = \exp\left[i(\theta/2\pi)\int_{\Sigma}A_{\mu\nu\lambda}dx_{\mu}dx_{\nu}dx_{\lambda}\right]
\end{equation}
This is the analog of a Wilson loop in 2D $U(1)$ in the sense that, if $\Sigma$ is a closed 3-surface that forms the 
boundary of a 4-dimensional volume $V$, inserting the Wilson bag
factor (\ref{eq:WB}) in the gauge field path integral is equivalent to including a $\theta$-term in the gauge
action over the 4-volume $V$. The discussion of what happens as we vary $\theta$ from 0 to $2\pi$ is also 
completely analogous to the screening of the 2D Wilson loop. For a fractional bag charge $\theta/2\pi$, 
with $0<\theta<2\pi$, the vacuum inside the bag will
have a higher energy than the $\theta=0$ vacuum outside. The expectation of the 
Wilson bag integral thus satisfies a 4-volume law analogous to the area law for the Wilson loop in 2D,
\begin{equation} 
 \label{eq:bag}
\langle B(\Sigma)\rangle \sim \exp\left[-(E(\theta)-E(0))V\right]
\end{equation}
There will thus be ``bag confinement,'' 
a confining force between the walls of a fractionally charged bag. 
At $\theta=\pi$, the vacuum inside the bag will 
undergo a nonanalytic shift corresponding to the transition
between two adjacent $k$-vacua, with parameters $\theta$ and $\theta-2\pi$ respectively. Finally, as we increase $\theta$
from $\pi$ to $2\pi$, the force between the bag walls decreases. For a unit-charged bag,
$\theta=2\pi$ inside the bag, and the confining force between bag walls disappears. The topological
charge is the curl of the Chern-Simons tensor, so for a uniform $A_{\mu\nu\lambda}$ which is nonzero on a 
flat bag surface, the topological charge distribution is a dipole layer consisting of thin, coherent positive and 
negative 3-dimensional layers on either side of the bag surface. Like the Wilson line excitations in the 
$CP^{N-1}$ models, screened unit-charged Wilson bags provide a reasonable model for interpreting
the topological charge structure observed in lattice configurations.

\section{Large N Chiral Lagrangians}

    As first emphasized by Witten \cite{Witten79}, considerations of large $N$ chiral symmetry also
point to a picture of the vacuum in $SU(N)$ gauge theories consisting of discrete quasi-vacua separated by domain walls. 
For large $N$ the size of the chiral $U(1)$ anomaly shrinks like $1/N$, which justifies treating the flavor singlet $\eta'$ meson as a would-be Goldstone boson and including it in the chiral Lagrangian.
The effect of the anomaly is incorporated in the chiral Lagrangian in the form of an $\eta'$ mass
term constructed from the $U(1)$ phase of the chiral field,
\begin{equation}
\label{eq:anom}
{\cal L}_{anom} = \frac{const.}{N}(-i\ln \;Det\; U)^2
\end{equation}
where $U$ is the $U(3)\times U(3)$ chiral field.
The $\eta'$ mass term arising from the anomaly  
has a different structure than a meson mass term coming from explicit
chiral symmetry breaking by quark masses, which has the form
\begin{equation}
\label{eq:qm}
{\cal L}_{qm} \propto Tr\left(\chi^{\dagger}U + h.c\right)
\end{equation}
where $\chi$ is the quark mass matrix.
In terms of the $U(1)$ phase, 
\begin{equation}
Det\;U = e^{i\eta}
\end{equation}
the anomaly term (\ref{eq:anom}) is a purely quadratic mass term $\propto \eta^2$, 
unlike the quark mass term (\ref{eq:qm}) which includes higher order multi-pion interactions
and is a single valued function of $U$.
The form (\ref{eq:anom}) for the anomaly term is dictated by large-N arguments and/or OZI phenomenology, 
in which multiple-hairpin vertices are suppressed. 

In fact, it is the multivaluedness of the logarithm in (\ref{eq:anom}) which leads to the appearance
of multple k-vacua and domain walls. To illustrate the point in it's simplest form, consider
the case of 1-flavor QCD, where the chiral field $U$ reduces to a single $U(1)$ phase, the would-be
Goldstone field,
\begin{equation}
U\rightarrow e^{i\eta}
\end{equation}
Now we consider the effective potential for the phase field $\eta$, including both the quark mass term
and the anomaly term. The potential is of the form
\begin{equation}
\label{eq:potential}
{\cal V} = {\cal V}_0(\cos\eta) +\frac{m_0^2}{N}\eta^2
\end{equation}
where $m_0^2$ is of order $\Lambda_{QCD}$. We assume that the potential term ${\cal V}_0$
has a minimum at $\eta=0$ and is periodic in $\eta\rightarrow\eta+2\pi$.
In the large N limit the anomaly term can be treated as small, and the potential ${\cal V}$ has
many nearly-degenerate minima, where the field $\eta$ differs by integer multiples of $2\pi$.
But the chiral anomaly allows us to equate a chiral $U(1)$ phase rotation with a shift of the
$\theta$ parameter. This leads to the conclusion that the quasi-vacua identified from the potential (\ref{eq:potential}) are 
k-vacua with $\theta$ parameters differing by integer multiples of $2\pi$. 
Thus large-N chiral Lagrangian considerations lead us to a picture of 
$\theta$-dependence with Wilson bags separating multiple, nearly degenerate k-vacuum
states characterized by effective local values of $\theta$ which differ by integer multiples
of $2\pi$. Although a chiral Lagrangian framework was invoked to arrive at this picture, it is
reasonable to conclude that the picture applies even to pure-glue QCD without quarks.
The quark only serves as a probe of the topological structure of the gauge field via it's chiral phase.

\section{Theta Dependence in Yang-Mills Theory from String/Gauge Holography}

A profound new source of intuition into the long rang structure of 4-dimensional gauge
theories has emerged over the last decade in the framework of AdS/CFT string/gauge duality. 
As Witten showed \cite{Witten98}, the string/gauge correspondence has particularly interesting 
implications for the structure of topological charge fluctuations in QCD. It 
nicely confirms the k-vacuum/domain wall scenario arrived at in the earlier work that I discussed
in the last two sections. In fact QCD topological charge provides a particularly direct
window on the stringy aspects of gauge theory. 
This is mainly due to the fact that, in the string/gauge correspondence, topological charge in 
gauge theory is dual to Ramond-Ramond (RR) charge in type IIA string theory. 
RR charge is of fundamental importance in string theory. It is similar to magnetic 
charge in electromagnetism, in that it is a solitonic charge which is not carried by ordinary string states. 
The fundamental discovery by Polchinski \cite{Polchinski} that D-branes carry Ramond-Ramond charge ushered in a new era in 
string theory where D-branes assumed a central role in the theoretical infrastructure.
D-branes were originally conceived for reasons associated with T-duality, which suggested considering
open strings with their ends attached to subdimensional hyperplanes
with Dirichlet boundary conditions. But it was soon realized that the hyperplanes so defined were actual
physical objects that carried energy density, were flexible, and could support local oscillations.
The low energy world-volume theory which describes the small oscillations of a D-brane is typically
a supersymmetric gauge theory. Much of the technology for studying the connections between string
theory and gauge theory is based on various ``brane constructions'' obtained by considering
string theory in the presence of one or more branes (usually, but not always, flat, parallel
or superimposed, and filling some $(D+1)$-dimensional subspace of 10-dimensional space.). 
By superimposing $N$ D-branes
in the same subspace, we obtain a world volume theory with a $U(N)$ gauge group, where the gauge
bosons correspond to short pieces of string connecting pairs of D-branes. With the interpretation
of the $N$ of the gauge group as the number of branes, an interesting thing typically happens in 
the large N limit. The gravitational mass of the N D-branes becomes large enough to form a 
black hole in the dimensions transverse to the branes. From studies of string theory near the 
horizon of a black hole, Maldacena \cite{Maldacena} was led to his famous conjecture that 4-dimensional
${\cal N}=4$ supersymmetric $SU(N)$ gauge theory is not simply the low energy limit of a string
theory, but is in fact ``holographically'' equivalent to type IIB string theory in the space 
$AdS_5\times S_5$. The idea that gauge theory in 4 dimensions is a holographic representation
of a higher-dimensional string theory is not only intriguing but 
has already gone a long way toward illuminating some of the partial understandings of QCD that have been around for many years. In the 
gravitational context, the idea of holography has its origin in the observation by Beckenstein and
Hawking that the entropy of a black hole is proportional not to it's volume but to the surface
area of it's horizon. It is as if everything that went on inside the black hole was uniquely
encoded on it's horizon. In the AdS/CFT correspondence, the behavior of weakly coupled string
theory in the 5-dimensional $AdS$ space maps holographically to a strongly coupled supersymmetric
gauge theory on the 4-dimensional boundary of that space.

In the equivalence conjectured by Maldacena, the 4-dimensional gauge theory that is equivalent
to IIB string theory in $AdS_5\times S_5$ is a conformally invariant field theory, ${\cal N}=4$
supersymmetric Yang Mills. In Anti-deSitter space, conformal symmetry of the corresponding gauge
theory is generic. However, Witten showed
that, with an appropriate arrangement of D-branes and boundary conditions, it is possible to
establish a similar holographic equivalence between string theory and ordinary, nonsupersymmetric,
asymptotically free gauge theory in 4 dimensions. Gravitationally speaking, Witten's construction
replaces the pure $AdS$ space of Maldacena with a Schwarzchild black hole metric, 
(which arises naturally in the large-N brane construction). 
It is possible to view Witten's construction
as an AdS/CFT correspondence, but only by going to 11 dimensional M-theory. 
An equivalent but somewhat more direct path to 4-dimensional QCD  
is a brane construction \cite{Witten98_brane,Witten98} which begins with a stack of $N$
4-branes in 10-dimensional IIA string theory on the spacetime manifold $R_4\times S_1\times R_5$.
The 4-branes (which have 5 spacetime dimensions) fill the subspace $R_4\times S_1$, i.e. they are
wrapped around the compact dimension, with supersymmetry breaking boundary conditions imposed
on the $S_1$. In the dimensions transverse to the 4-branes, this induces a 5-dimensional black hole
metric, and the global geometry changes from $R_4\times S_1\times R_5$ to $R_4\times D\times S_4$,
where $D$ is a 2-dimensional disk with a Schwarzchild singularity at its center. The $R_4$
is interpreted as 4D spacetime. For our purposes,
the $S_4$ plays an essentially passive role except as a place to wrap 6-branes. In the discussion
of holography and QCD topological charge, the crucial concept provided by Witten's brane construction
is the disk $D$ attached to each point in 4-dimensional spacetime. To understand how $\theta$-dependence of QCD
arises in the string theory context, we note that the world-volume
theory on the 4-branes is actually a 5-dimensional gauge theory on $R_4\times S_1$. Being odd-dimensional,
this world volume theory generically includes a 5-dimensional Chern-Simons term which has the 
form
\begin{equation}
{\cal L}_{CS_5} = \frac{1}{8\pi^2} a\wedge Tr (F\wedge F)
\end{equation}
where $Tr(F\wedge F)$ is the topological charge density in the 4-dimensional theory, and $a$ is the
Ramond-Ramond $U(1)$ field around the compact $S_1$. In the limit of small compactification
radius, the 5D Chern-Simons term reduces (at least locally) to a 4D theta term
\begin{equation}
{\cal L}_{CS_5} \rightarrow \frac{\theta}{16\pi^2} F\wedge F
\end{equation}
where the value of $\theta$ is given by the line integral of the RR $U(1)$ field around the compact
dimension,
\begin{equation}
\label{eq:thetaloop}
\theta = \oint_{S_1} a_5 dx_5
\end{equation}
Here $S_1$ is around the perimeter of the disk $D$. Thus $\theta$ is proportional to the amount 
of Ramond-Ramond flux threaded through the singularity at the center of the disk:
\begin{equation}
\theta = \int_D f_{\mu\nu}dx_{\mu}dx_{\nu}
\end{equation}
where 
\begin{equation}
f_{\mu\nu}=\partial_{\mu}a_{\nu}-\partial_{\nu}a_{\mu}
\end{equation}

In the holographic framework, the radius of the compact $S_1$ serves as an ultraviolet
cutoff, rather analogous to the lattice spacing in a lattice formulation. In the limit of
small radius, the value of $\theta$ defined by the line integral (\ref{eq:thetaloop}) will
approach a spacetime constant $\theta$ parameter, but only mod $2\pi$. Different regions
of space may be in different k-vacua, i.e. have values of $\theta$ which differ by multiples
of $2\pi$ and thus are separated by domain walls. The string theory interpretation of
$\theta$ as the Wilson loop of the RR field around the compact dimension thus leads naturally
to the existence of k-vacua. It also points to the correct candidate for the 
string theory analog of a gauge theory Wilson bag.
In IIA string theory, a special role is played by D6-branes and I will now argue
that the topological charge membranes seen on the lattice can be interpreted as the 
holographic image of D6-branes which are wrapped around the compact $S_4$ and therefore
appear as 2-branes or membranes in 3+1 dimensions. I will show that the defining
property of the Wilson bag, namely that the value of $\theta$ jumps by 
$\pm 2\pi$ when crossing the surface, is in fact nothing but the statement of quantization
of Ramond-Ramond charge on a D6-brane. A basic property of the D6-brane is that it carries 
a quantized amount of RR charge, where this quantization can be demonstrated by a generalization of
Dirac's magnetic monopole construction. In general, Dirac's argument involves integrating the
magnetic flux over a 2-surface which surrounds or ``links with'' the object that carries the
magnetic charge. Two objects of dimensionality $d_1$ and $d_2$ which link with each other in 
D spatial dimensions satisfy $d_1+d_2=D-1$. So in $D=9$ spatial dimensions, the spatial
dimensionality of an object which can be surrounded by a 2-surface is $d=6$, i.e. a D6-brane.

\begin{figure}
\vspace*{2.0cm}
\includegraphics{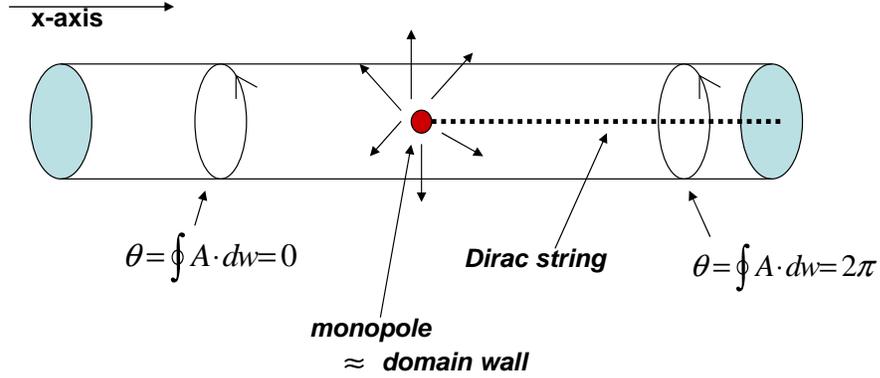}
\vspace{6.5cm}
\caption{Holographic view of a domain wall between k-vacua in QCD. Plot is at a fixed time, with
the long axis of the cylinder representing the spatial coordinate transverse to the domain wall
at a fixed time.
$w$ is the coordinate on the circle around the compact dimension.}
\label{fig:laughlin}
\end{figure}

A pictorial representation of this situation is shown in Fig. \ref{fig:laughlin}. Here the 
solid cylinder represents the disk $D$ at every point along a spatial axis transverse to 
the domain wall, (e.g. we put the domain wall in the $y$-$z$ plane, and the long axis of the
cylinder is the x-axis.) This dimensionally reduced picture makes it obvious that the 
quantized step in $\theta$ across a domain wall comes about in the string theory 
by a Dirac-type quantization of RR charge. The surface surrounding the 6-brane consists of
two disks located on opposite sides of the domain wall (slightly distorted and joined around the
outer edge to form a closed surface). The difference between the amount
of RR flux going through the two disks, i.e. the step in $\theta$, is quantized.

The picture in Fig. \ref{fig:laughlin} serves equally well to represent a domain wall in the
2D $CP^{n-1}$ models. At a fixed time, a domain wall is located at a point on the spatial x-axis.
Again, the domain wall is analogous to a magnetic monopole
with a quantized Dirac string coming out of it.

\section{Small N, Large N, and Melting Instantons}

As Witten convincingly argued \cite{Witten79}, a picture of the QCD vacuum based on instantons
is in fundamental conflict with expected properties of the large $N$ limit. 
In fact, in any simple instanton model, the mass of the $\eta'$ meson is exponentially
suppressed, $\propto \exp(-const.\times N)$. But if chiral symmetry breaking occurs as expected
in the large $N$ limit, arguments associated with the Witten-Veneziano relation imply that the
$\eta'$ mass$^2$ is $\propto 1/N$. This and other problems with the instanton picture led Witten
to suggest that, at sufficiently large $N$, instantons would ``melt'' due to large field fluctuations
associated with the confining vacuum. As we have seen, the idea that topological charge
comes in the form of codimension 1 membranes is in a sense the large-$N$ alternative to the
instanton picture. The question of whether $N=3$ is large enough for large $N$ arguments 
to apply to 4-dimensional gauge theory is clearly a central issue. 
A recent study of $N$-dependence of topological structure in the $CP^{N-1}$
models \cite{Lian06} provides an interesting perspective on the phenomenon of melting instantons.
It turns out that in these models there is a fairly precise instanton melting point
$N=N_c$ below which instantons dominate the topological charge distribution on the lattice, and topological
susceptibility is reasonably well-described by a dilute instanton gas calculation.
Above the melting point, instantons disappear and the TC distribution is dominated
by coherent codimension 1 surfaces.

For $CP^{N-1}$ with the lattice action used in \cite{Lian06} the instanton melting point is $N_c\approx 3.7$.
Thus $CP^1$ and $CP^2$ are below the melting point, and $CP^3$ ($N=4$) is slightly above it.
Using the overlap-based $q(x)$, the small instantons which dominate $CP^1$ and $CP^2$ are
quite easily seen, and, at larger values of $\beta$, integer changes of the global 
charge $Q$ are invariably accompanied by the appearance or disappearance of an identifiable
instanton. (Interestingly, the instantons are not visible if the ultralocal plaquette-based 
definition of $q(x)$ is used.)
For reasons first discussed by Luscher \cite{Luscher82} the instantons 
in these models have radii of order lattice spacing. They have zero radius in the continuum
limit and result in anomalous scaling of the topological susceptibility. The instanton
melting point can be estimated in the dilute gas approximation by measuring the action $\epsilon$ of
a single small instanton on the lattice. Numerically, this is found to be approximately
\begin{equation}
\epsilon \approx \frac{N}{2}\times 6.74\ldots
\end{equation}
For $N=1$ and $2$, the instanton contribution falls more slowly than $\mu^2$
in the large $\beta$ limit ($\mu=$ mass gap). This results in a divergent topological susceptibility in
the continuum limit. On the other hand, if $\epsilon>4\pi$ the instanton contribution to
$\chi_t$ falls off more rapidly than $\mu^2$ and becomes negligible compared to scaling
contributions which are $\propto \mu^2$
Thus there is a sharp transition at $N=N_c = 8\pi/6.74 \approx 3.7$. For $N<N_c$, instantons dominate
and for $N>N_c$ they disappear. The instanton melting point $N_c\approx 3.7$, estimated from the dilute gas 
calculation, agrees very nicely with the direct results of the Monte Carlo calculations,
which give a divergent $\chi_t$ for $CP^1$ and $CP^2$ but a $\chi_t$ which scales properly
for $CP^3$ and higher \cite{Lian06}. 

Using the $CP^{N-1}$ analogy, it is possible to at least crudely estimate 
the instanton melting point in 4D $SU(N)$ gauge theory.
First we note that the action of a small instanton, and hence the estimated value of $N_c$
will change with different lattice actions. So it may be possible to improve
the action in a way that would increase the value of $\epsilon$ and thereby lower the value of 
$N_c$. But Luscher has argued \cite{Luscher82} that the lattice effect will always lower the
action of a small instanton relative to its continuum value of $\epsilon_c =\frac{N}{2}\times 4\pi$
So there is a lower bound on the instanton melting point of $N_c>2$. (In particular, this means 
that small instantons cannot be eliminated from $CP^1$ by improving the action.)
The lower-bound estimate of $N_c=2$ can be
obtained without reference to any particular lattice action. $N_c=2$ corresponds to the
``tipping point'' of the integration over instanton size in a semiclassical instanton calculation.
For $N<N_c$ the integral diverges at the small instanton end, while for $N>N_c$, it diverges
for large instantons. For 4-dimensional $SU(N)$ gauge theory, the integral over instanton size
behaves (to lowest order in the renormalization group beta function) like
\begin{equation}
\int \frac{d\rho}{\rho^5} \;\rho^{11 N/3}
\end{equation}
which has it's tipping point at
\begin{equation}
\label{eq:melt}
N_c=\frac{12}{11}
\end{equation}
Real QCD at $N=3$ is well above this estimate of the instanton melting point, but (\ref{eq:melt})
is only a lower bound. For $CP^{N-1}$, the actual melting point (for the simplest action)
is nearly twice as large as the lower bound (3.7 vs. 2). In any case, the direct Monte Carlo evidence
from $SU(3)$ gauge theory clearly favors the large $N$ scenario of codimension 1 membranes
rather than an instanton dominated vacuum. 

The phenomenon of melting instantons in the $CP^{N-1}$ models has an amusing interpretation
in the framework of string/gauge holography. By analogy with Witten's 4-brane construction, we
can imagine that the $\theta$ term in 2D $CP^{N-1}$ arises from a compactified 3-dimensional 
Chern-Simons term of the form
\begin{equation}
{\cal L}_{CS} = i\epsilon^{abc}A_a \partial_b A_c
\end{equation}
Here, $a, b, c$ run from 1 to 3.
Let us denote the original spacetime dimensions by 1 and 2, and the compactified dimension by 3.
Then in the limit of small radius of compactification, the Chern-Simons term reduces to a theta term,
\begin{equation}
{\cal L}_{CS} \rightarrow i\frac{\theta}{2\pi} \epsilon^{\mu\nu}\partial_{\mu}A_{\nu}=i\theta q(x)
\end{equation}
where $\mu,\nu = 1,2$, and
\begin{equation}
\label{eq:theta}
\theta = \oint A_3 dx_3
\end{equation}
From this 3-dimensional framework, a small instanton in $CP^1$ or $CP^2$ can be interpreted as a charged particle coupled
to the gauge field $A_3$ which has a world line which is wrapped around the compact direction in a closed loop, 
and is pointlike in the 1-2 plane. On the other hand,
we may integrate by parts and write,
\begin{equation}
{\cal L}_{CS} = -\frac{i}{2\pi} \epsilon_{\mu\nu}(\partial_{\mu}\theta)A_{\nu} \equiv J_{\nu}A_{\nu}
\end{equation}
where
\begin{equation}
\label{eq:dwcurrent}
J_{\nu} \equiv \frac{1}{2\pi}\epsilon_{\mu\nu}\partial_{\mu}\theta
\end{equation}
In this way of writing the CS term, the current $J_{\mu}$ couples to the gauge field in the 1-2 plane. 
In the limit of small compactification radius, the quantity $\theta$ defined by (\ref{eq:theta}) reduces
to the constant theta parameter of the 2D theory mod $2\pi k$.  
Different k-vacua are separated by domain walls, and the current $J_{\mu}$ is an ``edge current'' which
is nonvanishing along these domain walls. 

From this perspective, we could interpret the melting of an instanton
as an unwinding of its world line from the compact dimension, with the domain walls of the large $N$ models being 
the remnants of melted or unwound instantons. The basic instability that causes instanton melting is 
that, for sufficiently large N, they are unstable toward expanding
in size. However, they cannot just become large instantons, because that would violate the negativity of the correlator.
Instead they expand like a smoke ring and become Wilson line excitations. Most of the time this ring of positive
$q$ will be screened by an antiinstanton emerging from the lattice and becoming a concentric ring of negative
$q$. forming a dipole layer.
This visualization of a melting instanton has the obvious generalization to 4-dimensional QCD, with
a small instanton expanding to form a hollow bubble whose surface is a Wilson bag.

\section{Conclusions}

The Monte Carlo results showing the presence of extended, coherent 3-dimensional topological charge 
membranes in lattice QCD configurations could have far-reaching implications. Detailed studies of low
Dirac eigenmodes and their relation to the topological charge membranes should clarify the role of 
these membranes in spontaneous chiral symmetry breaking. 
It is likely that the chiral condensate arises from low eigenmodes
associated with these membranes. The chiral anomaly and $\eta$' mass insertion
suggest a central role of the membranes in inducing 
quark-antiquark pair creation and annihilation processes in the QCD vacuum. The D-brane picture of the
QCD vacuum could provide a better
understanding of the OZI rule, in particular the fact that $q\bar{q}$ annihilation within a meson is highly suppressed 
except in the scalar and pseudoscalar channels. The interpretation of the coherent topological charge
sheets as Dbranes also suggests a possible underlying role of supersymmetry in the dynamics of light quarks. 
Such a connection has already been found in the ``supersymmetry relics'' 
discussed recently by Armoni, et al \cite{Armoni}. These predictions are based on a large-N equivalence 
between ${\cal N}=1$ SUSY gauge theory and an orientifold projected theory which, for the
case $N=3$ is ordinary non-SUSY 1-flavor QCD. As shown
in the poster of Patrick Keith-Hynes at this Conference,\cite{Patrick} the SUSY relic prediction of approximate degeneracy between scalar and 
pseudoscalar flavor-singlet mesons depends crucially on the mass shifts induced by $q\bar{q}$ annihilation 
(hairpin) diagrams in the scalar and pseudoscalar channels. The interplay between light quark dynamics and
topological charge membranes in QCD is an interesting area for both theoretical and numerical studies.
Finally, the interesting question arises whether the D-brane vacuum I have suggested in this talk can explain
confinement. This is at least a reasonable possibility, since the presence of topological charge
membranes should disorder the vacuum at long distances. A large Wilson loop would have to 
go through many membranes, which would disorder the color phases around the loop and lead to an area law.
Intuition from the string theory side should be useful in addressing these issues.

This work was supported in part by the Department of Energy under grant DE-FG02-97ER41027.


\begin{thebibliography}{99}

\bibitem{Hasenfratz97} P. Hasenfratz, V. Laliena, F. Niedermayer, Phys. Lett. B427, 125 (1998).

\bibitem{Neuberger} H. Neuberger, Phys. Rev. Lett.81, 4060-4062 (1998).

\bibitem{Horvath03} I. Horvath et al, Phys. Rev. D68, 114505 (2003).

\bibitem{Lian06} Y. Lian and H. B. Thacker, hep-lat/0607026.

\bibitem{Horvath_corr} I. Horv\'ath et al., Phys. Lett. B617: 21 (2005).

\bibitem{Ilgenfritz} E.-M. Ilgenfritz et al., Nucl. Phys. Proc. Suppl. 153: 328 (2006). 

\bibitem{MILC} MILC Collaboration:C. Aubin et al,, Nucl. Phys. Proc. Suppl. 140: 626 (2005);
               C. Bernard et al., PoS LAT2005: 299 (2006).

\bibitem{Diakonov} D.I.~Diakonov, V.Y.~Petrov, Nucl.~Phys.~{\bf B272}, 457 (1986).

\bibitem{Ahmad05} S. Ahmad, J. T. Lenaghan, H. B. Thacker, Phys. Rev. D72: 114511 (2005).

\bibitem{Luscher78} M. Luscher, Phys. Lett. B78, 465 (1978).

\bibitem{Witten79} E. Witten, Nucl. Phys. B149, 285 (1979).

\bibitem{Witten80} E. Witten, Ann. Phys. 128: 363 (1980).

\bibitem{Coleman76} S. Coleman, Ann. Phys. 101: 239 (1976).

\bibitem{Witten98} E. Witten, Phys. Rev. D 81, 2862 (1998).

\bibitem{Polchinski} J. Polchinski, Phys. Rev. Lett. 75, 4724-4727 (1995).

\bibitem{Maldacena}
  For a review, see O.~Aharony, S.~S.~Gubser, J.~M.~Maldacena, H.~Ooguri and Y.~Oz,
  Phys.\ Rept.\  {\bf 323}, 183 (2000).

\bibitem{Witten98_brane} E. Witten, Adv. Theor. Math. Phys. 2: 505 (1998).

\bibitem{Luscher82} M.~L\"{u}scher, Nucl.\ Phys.\ B200:61 (1982).

\bibitem{Armoni} A. Armoni, M. Shifman, and G. Veneziano, Phys. Rev. Lett. 91: 191601 (2003).

\bibitem{Patrick} P, Keith-Hynes and H. B. Thacker (these proceedings).

\end{thebibliography}
\end{document}